\documentclass[prl,aps,amsmath, twocolumn, superscriptaddress]{revtex4}

\usepackage{graphicx, amsmath,amssymb}
\usepackage[latin1,applemac]{inputenc} 

\begin{document}

\title{Probing physical properties of confined fluids within
individual nanobubbles}

\author{D. Taverna}
\author{M. Kociak}
\author{O. St\'{e}phan}
\author{C. Colliex}
\affiliation{Laboratoire de Physique Solides, CNRS, UMR8502,
Universit\'{e} Paris-Sud, Orsay, France}

\author{A. Fabre}
\affiliation{C.E.A. Centre d' Etudes de Valduc,  Is-sur-Tille, France}

\author{E. Finot}
\affiliation{ Laboratoire de Physique, UMR CNRS 5027,Universit\'{e} de Bourgogne,Dijon , France}

\author{B. D\'{e}camps}
\affiliation{Laboratoire de Chimie M\'{e}tallurgique des Terres Rares, UPR 209 du CNRS, Thiais, France}
\author{C. Colliex}
\affiliation{Laboratoire de Physique Solides, CNRS, UMR8502,
Universit\'{e} Paris-Sud, Orsay, France}

        \date{\today}

\begin{abstract}
Spatially resolved electron energy-loss spectroscopy (EELS) in a scanning transmission electron microscope (STEM)
has been used to investigate as fluidic phase in nanoubbles embedded in a metallic $Pd_{90}Pt_{10}$ matrix. Using the
$1s \rightarrow 2p$ excitation  of the He atoms, maps of the He distribution, in particular of its density an pressure
in bubbles of different diameter have been realized, thus providing an indication of the involved bubble formation mechanism.
However, the short-range Pauli repulsion mechanism between electrons on neighboring atoms seems insufficient
to interpret minute variations of the local  local measurements performed at the interface between the metal and the He bubble.
Simulations relying on the continuum dielectric model have show that these deviations could be
interpreted as an interfzce polarization effect on the He atomic transition, which should be accounted for when measuring
the densities within the smaller bubbles.
\end{abstract}\pacs{} \maketitle

Confined fluids in nanosized volumes constitute challenging
objects for both basic and technological aspects.
The investigation of the structural features and dynamics of nanojets
has given rise to spectacular experimental studies and theoretical simulations
\cite{Moseler2000}. Another ideal system is represented by gas confined in nanocavities.
It is the case of inert gas atoms coalescing as a fluid or a solid to fill nanocavities in metals, with spherical or faceted morphologies depending of the local  pressure.  In the case of Xe in
Al, an interfacial ordering has been demonstrated by high
resolution electron microscopy \cite{Donnelly2002}. These small gas-filled cavities therefore behave as high-pressure
cells, providing the boundary conditions for the evaluation of the
physical properties of encapsulated gases. A most challenging problem is the evaluation of gas density and pressure in such cavities.

Among the possible systems, He nanobubbles in metals have attracted the attention of many researchers,
because of their high technological interest in the aging of the mechanical properties of materials involved in nuclear reactors \cite{Lucas1984}.
Measurements averaging the
information over large populations of bubbles, the size
distribution of which being controlled by TEM, have first been
performed by NMR \cite{Abell1987} and by a combination of optical
absorption and electron energy-loss spectroscopy (EELS)
\cite{Rife1981}. The first of these studies has revealed a
solid-fluid transition at  250K for bubble pressures ranging from
6 to 11 GPa (i.e. He atomic densities from about 100 to 200
nm$^{-3}).$
The second study comparing UV absorption spectroscopy
on a synchrotron and high energy resolution EELS without
spatial resolution on He$^+$ implanted Al thin foils, have
identified the blue shift of the He $1s  \rightarrow 2p$
transition (with respect to its value of 21.218 eV for the free
atom) as a hint for evaluating the local pressure .
Theoretically, Lucas {\it et al.} \cite{Lucas1983} have confirmed
that this blue shift of the He K-line should be attributed to the
short-range Pauli repulsion between the electrons of neighboring He atoms.
Consequently, this effect should increase
linearly with the density of the He atoms in the high-pressure
fluid phase likely to exist in these nanosized bubbles. J\"ager {\it
et al.} \cite{Jager1983} have confirmed this linear relation
between the measured energy shifts $(\Delta E)$ and the average
bubble radii $(r),$ the
larger shift corresponding to the higher He density and
consequently to the smaller radii.

With the development of scanning transmission electron microscopy
(STEM) techniques, capable of measuring spatially resolved EELS
spectra for different positions of a sub-nm probe on the specimen,
new possibilities were offered to perform analysis on individual
nanobubbles \cite{McGibbon1991}.
The most comprehensive study up-to-date has been conducted by
Walsh  {\it et al.} \cite{Walsh2000}, who proposed a procedure for
estimating directly the helium density in a
single nanobubble.
However, this work did not take into account the influence
of interface excitations on the estimation of the internal density.
A more fundamental issue which has not been addressed in the case of helium bubbles is the potential occurrence of density
inhomogeneities close do the surface, due to the interaction between the
confined fluid and the matrix.
The investigation of such effects requires a refined characterization
at a sub-nanometer scale.
\begin{figure}[t]
\begin{center}
   \includegraphics[width=6 cm]{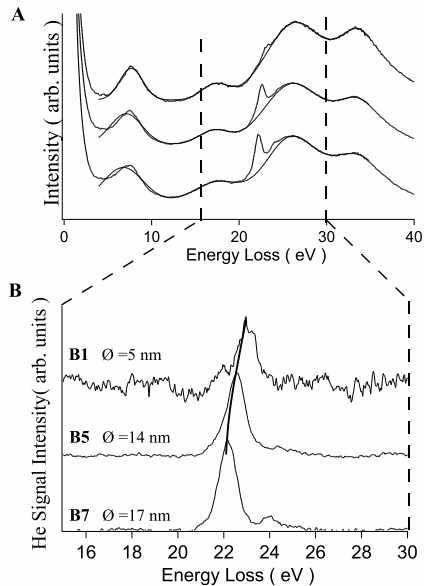}

\end{center}
  \caption{
          {\bf A:} EELS spectra acquired at the centre of
                   bubbles of different size (thick lines) and corresponding fit
                  of the Pd  plasmon (thin lines).
          {\bf B:} Subtracted He signal. The
                   shift of the He K-line for bubbles of different size is obvious.
                  }
\label{fig1}       
  \end{figure}

In this letter, we present a study of the physical parameters
(density, pressure, energy of the He K-line) defining the state of He inside nanobubbles,
by using spatially resolved EELS to map their variations at the nanometer scale.
The variations between bubbles of different size are in agreement
with the standard interpretation in the literature, while a refined description
is required for the evolution of the He signal within an individual bubble.
By using the continuum dielectric model, we show that the discrepancies can be explained
invoking an effect of surface polarization at the interface between the He and the metallic surface.
This leads to the necessity of a correction of the EELS estimation of the He density inside small bubbles.

\begin{figure}[t]
\begin{center}
 \includegraphics[width=8.5 cm]{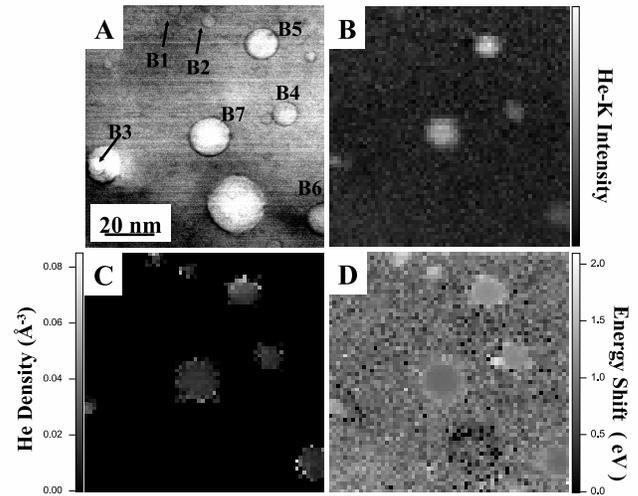}

\end{center}
  \caption{
           Maps extracted from a spectrum-image of a selected area of the sample.
           {\bf A:} Bright field image of the analyzed area.
                         Bubbles showing He signal are evidenced.
           {\bf B:} Helium chemical map.
           {\bf C:} Map of the He density inside the He filled bubbles.
           {\bf D:} Map of the energy shift of the He K-line. The reference energy is chosen as that of the atomic He (21.218 eV)
                }
\label{fig2}       
  \end{figure}

The results are issued from an 8-month aged
tritiated Pd$_{90}$Pt$_{10}$ alloy which exhibits
a largely dispersed population of voids  (from 2 to 20 nanometers in diameter).
The EELS measurements have been
performed in a VG STEM HB 501 with a field emission gun operated
at 100 kV and a homemade detection system formed by a Gatan 666 PEELS
spectrometer optically coupled with a CCD camera.
Spectrum-images made typically of 64$\times$64 spectra could then be acquired with the following
conditions : acquisition time of 200 ms per spectrum, probe of 0.7
nm with step increments of ranging from 1.5 to 0.5 nm.
Fig. \ref{fig1}A shows three EELS spectra corresponding to a selection of pixels at the centre
of three bubbles of different sizes (B1, B5 and B7) visible on
fig. \ref{fig2}A. These spectra correspond to positions where the electron beam
has crossed both the metallic matrix and the bubbles. They all
exhibit four major peaks (around 7, 17, 26 and 33 eV
respectively), which are attributed to the low energy loss
spectrum of the Pd alloy matrix. The sharper line between 22 and
23 eV is the signature of the He K edge.

In order to be more quantitative, the He signal of each spectrum has been isolated by
fitting the palladium contribution with 4 Gaussian curves (fig. \ref{fig1}B) .
 We can then identify any
change in position, width, total intensity and possible  occurrence of fine
structures or satellites on the He K-line, related  the
different bubbles. For each  probe position, the He
K-line intensity can be evaluated by integrating the signal over a
window of typically 4 eV and the results $(I_{He})$ are displayed as a
2D map (fig. \ref{fig2}B) of the localization of He atoms. It
must be noticed that not all the voids contain He atoms. The next
step is to transform the He K-line intensity map into a
cartography of the absolute estimated He density $n$
(expressed in atoms/nm$^{3}).$ This can be calculated from the relation \cite{Walsh2000}:
$n=I_{He}/(\sigma_{He} I_z d),$
where $\sigma_{He}$ is the cross section of the helium $1s \rightarrow
2p$  transition for the used experimental conditions
(see\cite{Walsh2000}  for calculation), $I_z$ is the integrated
intensity of the elastic peak, $d$ is the local thickness at the
pixel position of the analyzed He nano-volume.
 This parameter is the source
of highest uncertainties. We have tested several approaches, but
finally we estimated it experimentally as the complement to local
thickness measurements of the matrix.
The resulting density map is shown on fig. \ref{fig2}C.
\begin{figure}
\begin{center}
  \includegraphics[width=6.5 cm]{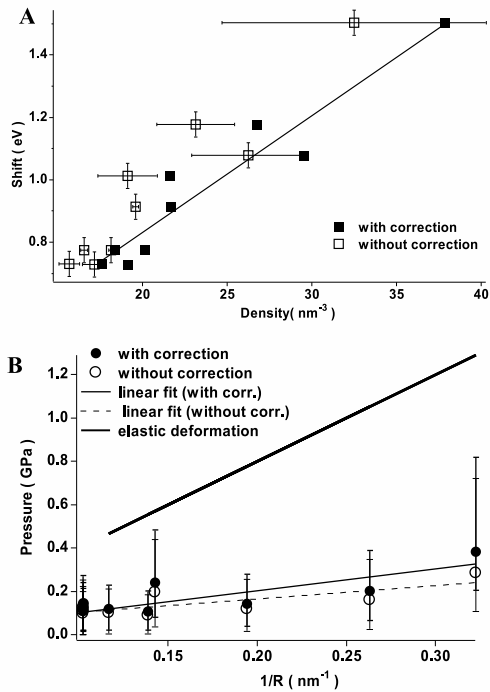}

\end{center}
  \caption{
          {\bf A:} Experimental relation between energy shift and measured
                   density. Empty squares represent uncorrected density value; linear fit law:
                   $\Delta E = ( 0.044\pm 0.007) n + ( 0.07 \pm 0.18).$
                   Filled square represent density value corrected by surface effects;
                   linear fit law:$\Delta E = ( 0.037\pm 0.004) n + ( 0.08 \pm 0.10).$
                   Error bars correspond the standard deviation calculated on the selection of pixels of the density map,
                   and therefore are large for small bubbles having a bad statistics.
          {\bf B:} Experimental relation between pressure and the inverse of the bubble radius.
                   Empty circles are deduced from uncorrected density values; linear fit law
                   $P = (0.6 \pm 0.2)\times1/R + (0.04\pm 0.05).$
                   Filled circles account for surface effects; linear fit law
                   $P = (1.0 \pm 0.4)\times1/R + (0.002\pm 0.083).$
                   The theoretical linear relation for elastic deformation of the Pd$_{90}$Pt$_{10}$
                   matrix is also displayed.
                   Error bars are estimated by calculating the variation of the equation of state in the density range
                   defined by the corresponding density error bars.
                   Error bars for corrected values (not shown) are identical to those for uncorrected values
                 }
\label{fig3}       
  \end{figure}

The mean helium density inside a bubble is estimated by averaging the
calculated values over a selection of pixels corresponding to
central positions. The results range from 15 to 35 He atoms per
nm$^3,$ the highest value been obtained for one of the smaller
bubble B1.
The energy shift, defined as the
difference between the measured peak position inside the
bubbles and the nominal K-line of atomic He
\cite{Lucas1983},  is mapped on fig. \ref{fig2}D and
varies from about 1 up to 2 eV.
In order to verify the predicted linear dependence of $\Delta E
(n)$, we have plotted in fig. \ref{fig3}A our results issued
from several spectrum-images (empty squares).
A satisfactory fit to a law  $\Delta E=C_n n+D$ can be obtained
with $C_n = (44\pm 7)\cdot10^{-3} eV\cdot nm^3 $ and D = $0.07 \pm
0.18 eV.$ The value of $C_n$ lies significantly higher than those
measured by J\"ager {\it et al.} \cite{Jager1983} and Walsh {\it et
al.} \cite{Walsh2000} but is close to that determined by McGibbon
\cite{McGibbon1991}.

Another relevant parameter is the internal pressure.
In fact, if the bubble deforms the matrix elastically, the radius
dependence of the bubble pressure is supposed to obey an inverse
proportionality law $P=2\gamma/r$ (where $\gamma$ is the surface energy).
Following the procedure indicated in \cite{Walsh2000}, we calculate the pressure from the
measured $n$ by using a semi-empirical equation of states (see supplementary materials and \cite{Trinkhaus1983,Walsh2000}).
The results are shown in fig. \ref{fig3}B (empty circles). The
pressure inside the bubbles is shown to increase roughly from 0.1
to 0.3 GPa (i.e. in a range well below the solid to liquid
transition pressure), when the diameter of the bubble decreases
from 17 to 5 nm. A reasonable value for the surface energy of the Pd$_{90}$Pt$_{10}$
alloy is  $\gamma=1.9 J m^{-2},$ to be compared to our experimental slope
$0.3 J m^{-2}.$ Then, the
bubbles seem to be under-pressured at the moment of our TEM observation.
\begin{figure}
\begin{center}
  \includegraphics[width=7 cm]{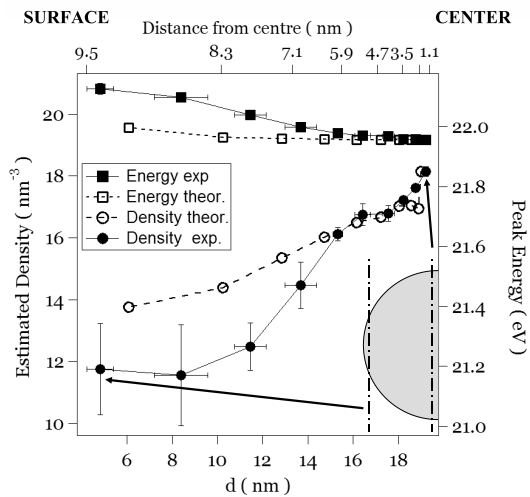}
\end{center}
  \caption{
            Experimental profiles of the estimated density ( filled circles) and of the blue-shift of
                            the He K-line (filled squares) as a function of the mean local thickness
                            $d.$ Data are extracted from a spectrum image of 40$\times$40 pixels (spatial sampling of 0.5 nm), acquired on a bubble of 19.5 nm diameter
                            For comparison, the corresponding simulated profiles of the He-K
                            line density  (empty circles) and energy
                            position (empty square)are also displayed.
                                }
\label{fig4}       
  \end{figure}

The spectrum-image technique offers the  extra possibility of
exploring any potential intra-inclusion spatial dependence. A varying
contrast seems visible within larger bubbles in fig. \ref{fig2}C and \ref{fig2}D, where the density
drops while the energy shift increases close to the surface of the
bubbles. In order to further investigate this behavior,
fig. \ref{fig4} shows experimental profiles of density
and shift as elaborated from a spectrum-image of a 19.5 nm bubble,
probed with a better lateral sampling of 0.5 nm.
Each point of the two profiles has been calculated
selecting annular regions of pixels corresponding to  the same
analyzed thickness, and fitting the He $1s \rightarrow 2p$
transition with a Gauss function to calculate the energy position and the intensity.
From the centre to the bubble surface a 37
percent  drop is observed for $n$ while the energy shift
increases by 0.17 eV, which is one order of magnitude smaller than
the shift between different bubbles. This anticorrelation is in
contradiction with the general tendency previously observed
between individual bubbles. Indeed, when only Pauli repulsion
between He atoms is taken into account, such a density drop
should lead to a 0.35 eV shift toward {\it lower} energies.
In order to evaluate the potential
occurrence of surface effects, we have performed EELS spectra
simulations,
by adapting  to the case of embedded spheres the continuum
dielectric model  which has proved its
efficiency for modeling local surface phenomena in nanosized
systems, such as single-walled nanotubes
\cite{odilePRB02,Taverna2003}. As an input for the simulation, we
used a lorentzian dielectric constant
corresponding to a He fluid of {\it constant} density  \cite{Lucas1983}. After simulation of
spectra for different local thicknesses $d$, the  procedure used to extract
$n$ and $\Delta E$ on experimental data is applied.
The resulting simulated profiles are compared to the experimental
ones in fig. \ref{fig4}B. Both $\Delta E$ and $n$ variations are reproduced but
underestimated.
Consequently, the major part of the effect can not be attributed to a real change in the density,
since the model assumes a constant one, but to the influence of surface excitations on the measurement.
We stress that the evidenced surface effect is not
due to an usual plasmon mode, because it does not correspond  to  a pole (resonance)
but to a maximum of the dielectric response of the sphere
$Im((\epsilon_{He}-\epsilon_{m})/(\epsilon_{He}+2\epsilon_{m}))$ (where $\epsilon_{He}$ and $\epsilon_{m}$ are
the dielectric constants of He and of the metallic matrix respectively).
An interface plasmon excitation is expected at a lower energy value (of the order of 7 eV),
and is rather unsensitive to the helium density.
However, the dielectric formalism
commonly used to model plasmon excitations furnishes reliable (similar)
interpretations for the effects of interface polarization on the atomic transition.
It is well known that such a formalism is very sensitive to the input dielectric
constants of both materials, and discrepancies between simulated and experimental
data can be partially explained by a lack accuracy of the lorentzian model adopted for He as well as
of the experimental data used for Pd.
Nevertheless, the energy shift of He K-line is explained by the
contribution of a surface ``mode'' which energy, for this particular system,
is slightly higher than that of the bulk He line (see supplementary materials).
The decrease of the estimated density can be related to a companion
effect known, for plasmons, as a boundary effect or ``Begrenzung''
effect \cite{geiger66}. This effect is commonly interpreted as a
modification in the probability to excite bulk modes due to the occurence of
surface excitation, and reveals itself as a negative contribution to the
intensity of the bulk He line,
the importance of which increases as the He thickness decreases.
We point out that this is the first time that this surface effect,
which has been thoroughly investigated for valence electron
excitations, is evidenced on an atomic-type excitation, using
EELS.

Therefore, beside its intrinsic fundamental interest, this
surface-induced decrease in the density estimated from atomic transition signal should
be taken into account in the study of the bubble formation mechanism.
We calculated a correction
coefficient $G$ to apply to experimental intensities in order to account for
surface effects in the estimation of $n$.
Such a coefficient is given by the ratio $G=I_{wos}/I_{tot},$ where $I_{wos}$ is the
He-K intensity simulated excluding surface contributions
(ideal case), and  $I_{tot}$ is the total simulated intensity(real case). The
resulting $\Delta E(n)$ corrected relation is displayed in fig.
\ref{fig3}A (filled squares).
The linear fit gives an estimation of the slope
C$_n$ decreased of 19 percent and closer to the values in the
literature. Even larger is the correction to relation between the pressure and the
inverse radius (filled circles in fig. \ref{fig3}B),
with a slope increased by a factor close to 2.
Nevertheless the comparison of the corrected data-set to the
linear relation characteristic of the elastic deformation regime
(also displayed in \ref{fig3}B) confirms that the bubbles are
under pressured.

In conclusion, the present analysis of the confined He fluidic phase,
at their interface with the embedding material, has evidenced an interface-induced effect
on the atomic-like spectral transition in He. Consequently, a reliable estimation of the helium internal density and
pressure requires a correction from this surface effect, especially in the case of small bubbles.
This effect can be of much broader interest,
and it should also be identified in quite different situations, such as those encountered
on semi-core loss edges (Hf-$O_{2,3}$) in dielectric thin films
\cite{couillard2007}.From another point of view,
the interpretation of residual discrepancies between experiments and simulations,
when exploring the influence of the distance from the interface, require further modeling,
accounting for changes in Pauli repulsion and in Van der Waals forces close to the interfaces.

We thanks L. Henrard, A. A. Lucas, Ph. Lambin and P. Loubeyre for fruitful discussions

\end{document}